\documentstyle[12pt]{article}
\begin{document}

\begin{titlepage}
\rightline{Jan 1999}
\rightline{UM-P-99/04}
\vskip 2cm
\centerline{\large \bf  
Have mirror stars been observed?}
\vskip 1.1cm
\centerline{R. Foot\footnote{Email address:
Foot@physics.unimelb.edu.au}}
\vskip .7cm
\centerline{{\it Research Centre for High Energy Physics}}
\centerline{{\it School of Physics}}
\centerline{{\it University of Melbourne}}
\centerline{{\it Parkville 3052 Australia}}
\vskip 2cm

\centerline{Abstract}
\vskip 1cm
\noindent
Observations by the MACHO collaboration suggest that
a significant proportion of the galactic halo
dark matter is in the form of compact objects with
typical masses $M\sim 0.5M_{\odot}$.  One of the current mysteries 
is the nature and origin of these objects. 
We suggest that these objects are
stars composed of mirror matter.
This interpretation provides a plausible explanation
for the inferred masses and abundance of the MACHO events.
We also comment on the possibility of inferring the
existence of mirror supernova's by detecting the
neutrino burst in existing underground detectors 
such as SuperKamiokande.

\end{titlepage}
\noindent
A long standing mystery is the nature of
the dark matter in the Universe.
In order to shed light on the nature of the dark
matter, the Australian - American MACHO collaboration
has been searching for Massive
Compact Halo objects (MACHO's) in the halo of our 
galaxy using the gravitational microlensing
technique suggested in Ref.\cite{pac}. 
By monitoring source stars
in the Large Magellenic Cloud the MACHO collaboration
have found 14 MACHO's events\cite{macho}.
These MACHO's have typical masses $M \sim 0.5M_{\odot}$
and they are estimated to make up a large fraction of
the galactic halo $f \sim 0.5$\cite{macho}.

So what are the MACHO's?
Over the last few years several conventional candidates
for the observed MACHO's have been discussed, such as
white dwarfs, brown dwarfs (i.e. Jupiter's), neutron stars, 
etc.  However all of these obvious candidates 
have either been excluded or appear to be disfavoured for
one reason or another
(for a recent discussion, see e.g. Ref.\cite{uu}).
For example, if the MACHO's are brown dwarfs, then
it is difficult to understand the estimated mass
of the MACHO events ($M \sim 0.5M_{\odot}$)
because brown dwarfs are much lighter ($\stackrel{<}{\sim}
0.08M_{\odot}$).  In view of the estimated masses of the MACHO's,
the most likely conventional interpretation is that
they are white dwarfs.
However if the MACHO events are white dwarfs
then it is very difficult to understand why
there are so many of them in the halo of our galaxy.
Furthermore there appears to be problems with overproduction
of heavy elements and overproduction
of light at high redshifts from the luminous stars which were the 
progenitors of the white dwarfs\cite{cs}.
Thus, instead of solving the dark matter mystery the
MACHO experiment has apparently deepened the mystery.

It is also apparently possible
that the MACHO's are not actually in the halo
of our galaxy but are stars in the LMC.
The situation will hopefully become better understood
as more studies are done.
If it turns out that the MACHO's (or at least
a significant proportion of them) are in the galactic
halo then it may be possible that the observed 
MACHO's are something more exotic. 
If this is the case what could they be?
This is a quite nontrivial question, because  
the usual exotic dark matter candidates, such 
as the hypothetical neutralino, would not clump together to 
form MACHO's. Are there any obvious particle 
physics dark matter candidates which
have the required properties to form MACHO's?

The purpose of this note is to suggest that
the observed MACHO events are mirror stars composed of mirror 
atoms (i.e. mirror baryons and mirror electrons).
\footnote{
Of course this is not the only possibility, 
examples of other possibilities
are that the MACHO's are primordial black holes, see
e.g. Refs\cite{macho,uu} and references there-in).}.
The existence of a set of mirror particles is 
well motivated from a particle physics point of view, since
these particles are predicted to exist
if parity is an unbroken symmetry of nature\cite{flv} (the
general idea was independently and earlier discussed 
a very long time ago by Lee and Yang\cite{ly} other
historical details can be obtained from Ref.\cite{P} and
references there-in).  The idea is 
that for each ordinary particle, such as the photon, electron, proton
and neutron, there is a corresponding mirror particle, of 
exactly the same mass as the ordinary particle. For example,
the mirror proton and the ordinary proton 
have exactly the same mass\footnote{
The mass degeneracy of ordinary and mirror matter
is only valid provided that the parity symmetry
is unbroken, which is the simplest and theoretically most
attractive possibility. For some other
possibilities, which invoke a mirror
sector where parity is broken spontaneously 
(rather than being unbroken), see Ref.\cite{other}.}.
Furthermore the mirror proton is stable for
the same reason that the ordinary proton
is stable, and that is, the interactions of the mirror
particles conserve a mirror baryon number.
The mirror particles are not produced
in Laboratory experiments just because they do not couple to any of
the ordinary particles. In the modern language of gauge
theories, the mirror particles are all singlets under 
the standard $G \equiv SU(3)\otimes SU(2)_L \otimes U(1)_Y$
gauge interactions. Instead the mirror
particles interact with a set of mirror gauge particles,
so that the gauge symmetry of the theory is doubled,
i.e. $G \otimes G$ (the ordinary particles are, of 
course, singlets under the mirror gauge symmetry).
Parity is conserved because the mirror particles experience
$V+A$ mirror weak interactions
and the ordinary particles experience the usual $V-A$ weak
interactions.  The only force common to both 
ordinary and mirror matter is gravity.

It is important to appreciate that compact 
objects like stars and planets would not be expected
to contain equal amounts of ordinary and mirror matter\cite{1}.
This is just because ordinary and mirror atoms cannot
collide with each other and dissipate energy to become
bound up in compact objects like stars. One naturally
expects the formation of stars composed primarily
of ordinary matter, and mirror stars composed primarily
of mirror matter\cite{1}.
Thus the segregation of ordinary and mirror matter
in the Universe can be nicely explained without
any additional assumptions
\footnote{An important issue
which we have not addressed is the distance
scale of this segregation. The interpretation
of the MACHO's as mirror stars would suggest
that this segregation distance is less
than or of the order of the inferred 
diameter of our galaxy (i.e. of order 30 kpc).
Whether or not this feature can be explained
in simple models of galaxy formation
in the early Universe is beyond the 
scope of the present paper. }.
In particular it is not necessary to assume that the mirror baryon
number of the Universe is much less than the ordinary baryon
number. Indeed it is most natural for these two quantities
to be comparable in magnitude. (Of course the origin
of the baryon number of the Universe is not understood
so no definite conclusions can be drawn). 
Note that in this respect mirror particles are quite different
to antiparticles, a Universe with equal amounts
of matter and anti-matter is known to be problematic. 

If mirror stars exist (i.e. stars composed of mirror atoms), 
then they would certainly appear to us as dark matter.
Such stars could only be observed by their gravitational
effects (unless they explode, as we will discuss later on). 
Note that the typical masses of mirror stars
should be similar to ordinary stars. This is just 
because the interactions of 
mirror atoms literally mirror those of ordinary atoms.
Thus providing that the Universe contains a significant
amount of mirror matter, the observed MACHO events can
be nicely explained. They are mirror stars,
being primarily composed of 
mirror hydrogen and mirror helium. 
The mirror stars would have masses of order of the solar
mass and this is nicely consistent with the 
observed MACHO 
events, which indicate that $M\sim 0.5M_{\odot}$.

Actually estimates of the contribution
of MACHO's to the mass density of the Universe
suggest that the mass density of MACHO's is of the
same order of magnitude as the mass density of
ordinary baryons\cite{uu}. This
feature may be plausibly explained by this mirror matter
interpretation of the data since it is quite natural
to expect that the mirror baryon number of the Universe
is comparable to the ordinary baryon number.

Note that the idea that mirror matter 
is the origin of some or all of the observed 
dark matter in the Universe has 
been proposed in earlier papers by a number of authors,
see e.g. Ref.\cite{1,cosmo}. The point of the present
paper is to point out that the gravitational effects of individual
mirror stars may have already been observed 
in the MACHO experiments.  
We have argued above that the rough features of the MACHO events
can be plausibly explained by this interpretation.
Of course, there is already strong evidence for the
existence of the mirror world coming from neutrino physics,
since one of the main predictions of these models\cite{P,P2} is that
each ordinary neutrino oscillates maximally into
its mirror partner if neutrinos have mass (this
is just a consequence of the unbroken parity symmetry).
This provides a very natural explanation for
the maximal mixing $\nu_\mu \to \nu_x$ ($\nu_e \to \nu_y$)
suggested by the atmospheric (solar neutrino experiments).

One immediate implication of this interpretation of the
observed MACHO's as mirror stars is that there
is a significant population of mirror stars in or near our
galaxy.  It is possible that mirror supernova also
occur. If this is the case then it may be possible to
detect the neutrino burst from such explosions. The reason
is that the mirror supernova should release a significant
amount of their energy into mirror electron neutrinos.
The mirror model interpretation\cite{P,P2} of the solar neutrino 
anomaly indicates that the electron neutrinos 
oscillate into mirror electron neutrinos with oscillation length
less than about 1 astronomical unit. 
Thus, we would expect half
of the mirror electron neutrinos from a mirror supernova explosion
to have oscillated into electron neutrinos.
If the mirror supernova explosion occurs in or near our galaxy
then these electron neutrinos should be detectable in various existing
underground neutrino experiments such as SuperKamiokande (just
as the neutrino burst from SN1987 was observed).
The obvious distinguishing signature of such explosions is 
that there will be no observed photon burst.
The discovery of such events would add further evidence
for the existence of a mirror world\footnote{
It should be pointed out that 
the properties of mirror supernova may not be
as similar to ordinary supernova as one might
naively expect.  (For example, the rate at which
mirror stars explode may be different to the rate at which
ordinary stars explode).  One reason for this is that the
primordial chemical compositions of ordinary and mirror stars
may be somewhat different.
This is possible because the ratio of primordial
mirror hydrogen to mirror helium may not be identical to
the ratio of ordinary hydrogen to helium even if
the baryon and mirror baryon numbers of the Universe are
equal in magnitude. 
Indeed big bang nucleosynthesis arguments suggest that
the temperature of the mirror particles in the early Universe
should be less than the ordinary particles. In this case
the ordinary light element abundances and the mirror light
element abundances would not be expected to be exactly same.}.
The typical energies of supernova neutrinos 
are in the range 10 to 25 MeV. It is tempting to speculate that
the small observed excess of solar neutrinos with energies
greater than about 13 MeV\cite{sk} may be due to electron neutrinos
emitted from distant mirror supernova.
However, there is no reason for their direction to be correlated with
the direction of the sun so that 
this explanation is probably not viable (although this
conclusion may depend on the details of the background
subtraction). 
Of course if the mirror supernova is close enough (i.e. in our galaxy
or in a nearby galaxy) then there should be several detected events in
a very short period of time (typically of order 10 seconds). The inferred
direction of the detected neutrino 
events should also be correlated. This should
be enough information to infer the existence of a mirror supernova even if
no photon burst is observed.

We conclude this paper with the
following general summary of
the implications of a mirror world.
There seems to be essentially three distinct ways
to test this idea\footnote{
There are two other possibilities which have
also been discussed in the literature,
i.e. photon-mirror photon mixing and higgs - mirror higgs 
mixing\cite{flv}. However, big bang nucleosynthesis arguments
suggest that such interactions are probably not observable
experimentally\cite{cg}\cite{lew}.}.
First, if neutrinos have mass then ordinary
and mirror neutrinos should maximally mix
with each other\cite{P}\cite{P2}. There is already strong 
experimental evidence that this mixing has already been
observed in the solar and atmospheric neutrino experiments
(the additional evidence for neutrino oscillations
obtained by the LSND collaboration is also compatible
with the mirror models\cite{P}).
This interpretation of these experiments will be
tested more rigorously in the near future as
more data is analysed and more experiments come on-line.
Second, the mirror sector will have important
implications for early Universe cosmology.
In general the predictions of the mirror model 
will be distinct from the standard model due
to the creation of lepton number asymmetries due
to ordinary - mirror neutrino oscillations\cite{asy}.
Early Universe cosmology will be stringently tested
in the future by the MAP and PLANCK experiments
and by improved estimations of the primordial
light element abundances.
Finally, the existence of mirror stars and perhaps
mirror galaxies is a natural consequence of
the existence of a mirror sector.
Such objects may help explain the inferred dark
matter of the Universe. 
Indeed, we have argued in this paper that
upto 14 mirror stars have already been `observed'
by the MACHO collaboration. We await
with interest for future experiments
and for a mirror star to explode
so that we can further test the existence of
a mirror world.

\vskip 0.3cm
\noindent
{\bf Note added}
\vskip 0.3cm
After submitting this paper, S. Blinnikov has informed me that
he has already discussed the possibility that the MACHO events are
mirror stars. See astro-ph/9801015. 
Also, after my paper was submitted, R. Mohapatra and V. Teplitz
(astro-ph/9902085) have also discussed MACHO's as mirror 
stars in the context of a model with mirror symmetry spontaneously
broken.

\vskip 0.4cm
\noindent
{\bf Acknowledgement}
\vskip 0.4cm
\noindent
The author thanks G. Filewood, N. Frenkel, H. Georgi and R. Volkas 
for a useful comment on this paper.  The author also thanks N. F. Bell
for explaining to me something about black holes which I 
should have known.  The author is an Australian Research Fellow.

\end{document}